\documentstyle[aps,epsfig,epsf]{revtex}
\begin{document}
\draft
\title{Low energy excitations in fermionic spin glasses:\\  
a quantum--dynamical image of Parisi symmetry breaking}
\author{R. Oppermann$^{1,3}$ and B. Rosenow$^{2}$}
\address{
$^1$ Service de physique th\'eorique, CE
Saclay, 
F--1191 Gif--sur--Yvette, {\it France}\\
$^2$ MPI f\"ur Kernphysik, D--69029 Heidelberg, {\it Federal Republic of 
Germany}\\ 
$^3$ Inst. f. Theor. Phys., Univ. W\"urzburg D--97074 
W\"urzburg, {\it F. R. Germany} }
\date{january 1998}
\maketitle
\begin{abstract}
We report large effects of Parisi replica permutation symmetry breaking (RPSB) 
on elementary excitations of fermionic systems with frustrated magnetic 
interactions. The electronic density of states is obtained exactly in the 
zero temperature limit for ($K=1$)--step RPSB together with exact relations 
for arbitrary breaking K, which lead to a new fermionic and dynamical Parisi 
solution at $K=\infty$. The Ward identity for charge conservation 
indicates RPSB--effects on the conductivity in metallic quantum spin glasses. 
This implies that RPSB is essential for any fermionic system showing spin 
glass sections within its phase diagram.
An astonishing similarity with a neural network problem is also observed. 
\end{abstract}
We present the first solution to the question whether and in which way 
Parisi replica permutation symmetry breaking (RPSB) 
and the related nonconstant part of the Parisi spin 
glass order parameter function $q(x)$ \cite{parisi,binyg,FiHe} 
are displayed in the low temperature many body theory of fermionic systems 
with frustrated Ising--interactions, emphasizing the $T=0$--limit in 
particular. The Parisi function $q(x)$, defined on the interval $0\leq 
x\leq 1$ is known as the apparently exact solution of the infinite range
classical spin glass. The x--dependence is known to be comparable with a
nontrivial time--dependence, induced by Glauber dynamics \cite{binyg},
of the spin autocorrelation function 
$<\sigma(\tau_x)\sigma(0)>$ (small x corresponding to large times). 
As shown by Parisi the function $q(x)$ assumes a 
plateau value within $x_1\leq x\leq 1$ and differs from this sofar
conventional single order parameter picture only within $0\leq x\leq 
x_1=O(T)$, where it decreases towards zero at $x=0$ in the absence of a 
magnetic field. The RPSB--effect appears to disappear with 
$T\rightarrow 0$. 
Nevertheless we find and report here a large $O(T^0)$--effect to persist in 
many 
important physical quantities of the fermionic Ising spin glass, which is 
a minimal quantum generalization of the classical Sherrington Kirkpatrick 
model. This includes replica--diagonal fermion 
Green's function and fermion density of states, where at any step K of RPSB 
the set of different order parameters is seen to determine the
quantum--dynamical behaviour of the fermion propagator and of vertex 
functions. These effects are complementary to and not in contradiction with
recent replica--symmetric descriptions of $T=0$ quantum spin glass transitions
\cite{sro}. Parisi--RPSB \cite{parisi} is seen to decide the qualitative 
and quantitative features of the low energy excitation spectrum.
While results are presented for an insulating model, the effect appears to be
rather model--independent and should hence be felt in transport 
properties of models with additional hopping hamiltonian for example.\\ 
The presence of spin glass phases within phase diagrams of interacting 
many--fermion systems such as High$T_c$superconductors, heavy fermion systems,
and semiconductors are nowadays recognized with increasing attention 
\cite{scalapino,mydosh,robr}. 
Many of their characteristic properties cannot be answered by considering 
these phases as isolated magnetic phenomena, which means that their common 
origin, their coexistence and competition with charge--related phenomena, and 
even far--reaching links into other fields of physics must be understood 
in terms of fermionic rather than spin space models.
The relationship between conductivity behaviour and magnetism, mainly
antiferromagnetism up to now, has acquired a prominent place in the conscience 
of theorists and experimentalists, due to the remarkable progress in the
field of strongly correlated fermion systems during recent years
\cite{vollhardt,georges}.\\
In this Letter we wish to provide results which evidence the fact 
that fermionic spin glasses also link closely glassy magnetic order and 
transport behaviour; further similarities between Hubbard model and the 
fermionic spin glass have been traced back to the particular role of the 
Onsager--Brout--Thomas reaction field \cite{binyg,vollhardt}
for all these systems, as can be observed by comparing Hubbard--CPA-- 
\cite{vollhardt} with fermionic TAP--equations \cite{robr}.\\
Spin-- and charge--excitation spectra of fermionic spin glasses must be 
evaluated in order to construct a meaningful many body theory.
This Letter focusses on the effect of Parisi replica permutation symmetry 
breaking (RPSB) on the single fermion density of states (DoS), 
hence on the fermionic Green's function, and, by virtue of the Ward
identity for charge conservation, also on vertex functions, thus on the entire
ensemble of quantities that provide the basis of many body theories for 
fermionic systems with frustrated interactions.\\
It is known that replica--diagonal quantities like the linear equilibrium 
susceptibility $\chi$ feel Parisi symmetry breaking even at $T=0$ 
\cite{parisi,binyg,FiHe} despite 
the fact that the nontrivial part of the Parisi function only lives on an 
interval of width T.\\ 
The susceptibility had been analysed by Parisi for the standard SK--model. 
He found a rapid convergence towards the exact result as the number of 
order parameters increased, this number being equal to $K+1$ in the SK--
and equal to $K+2$ in fermionic models. 
While the low temperature regime of the SK--model had not been of particular 
interest from the point of view of phase transition theory, it becomes highly
important for fermionic spin glasses, since the $T=0$--theory of excitation 
spectra plays a crucial role and, for the additional reason that some 
models exhibit quantum phase transitions along the $T=0$--axis. 
Parisi nevertheless analysed the low T regime \cite{parisi} of the classical
SK--model finding that K--step RPSB on one hand 
provided increasingly good approximations but failed to completely remove 
the negative entropy and the instability problem at low enough 
temperatures unless $K\rightarrow \infty$. \\
In this Letter the effect of one step RPSB ($K=1$) on the density of states is 
presented in detail, followed then by an analytical relation valid for all K, 
which allows to determine the type of excitation spectrum present in the 
full Parisi solution for the fermionic Ising spin glass.
Despite the fact that the regime of deviation from a replica--symmetric
spin glass order parameter is only of $O(T)$, we find that it has a large 
$O(T^0)$--effect on the fermion density of states, the one--particle--,
and  many--particle Greens functions at $T=0$. This density of states is 
derived as usual from the imaginary--time (disorder--averaged) fermion 
Green's function $[-<T_{\tau}[a(\tau)a^{\dagger}(0)]>]_{av}$, which is
one of the decisive quantum--dynamical elements of any many body 
theory of fermionic spin glasses. This illustrates that, unlike the usual
picture of a Parisi solution being just a static order parameter function,
the fermionic picture must include the qualitative extension to 
dynamical quantities. Those become drastically altered by the nontrivial 
part of the Parisi solution which is otherwise invisible at $T=0$, hence 
providing a {\it quantum--dynamical image of RPSB}.
We first consider a generalized Parisi solution of the infinite--range 
fermionic Ising spin glass model. Its grand canonical Hamiltonian 
\begin{equation}
{\cal{H}}=-\sum_{ij}J_{ij}\hat{\sigma}_i\hat{\sigma}_j-H\sum_i\hat{\sigma}_i
-\mu\sum_i(\hat{n}_{i\uparrow}+\hat{n}_{i\downarrow}),
\quad\hat{\sigma}_i\equiv\hat{n}_{i\uparrow}-\hat{n}_{i\downarrow},
\quad 
\hat{n}_{i\sigma}\equiv a^{\dagger}_{i\sigma}a_{i\sigma}, 
\nonumber
\end{equation}
with fermion operators $a, a^{\dagger}$ and represents a Fock space extension 
of the SK--model. The magnetic couplings $J_{ij}$ of this insulating model 
are independent gaussian variables with zero mean value. 
The chemical potential
controls the occupation of magnetic and nonmagnetic states, 
where the latter ones reduce the freezing temperature, and 
leads to remarkable effects in the tricritical phase diagram \cite{brro}.
We restrict our discussion to half--filling. The fermion Green's function
can be derived as 
${\cal{G}}=\frac{\delta}{\delta\bar{\eta}}\frac{\delta}{\delta\eta}ln\Xi$ 
from the generating functional (generalization from $K=1$--step RPSB, given 
here for the sake of simplicity, to arbitrary K is standard)
\begin{eqnarray}
& &\hspace{-.5cm}\Xi_n(\eta,\bar{\eta})=e^{-\frac{N}{4}\beta^2 J^2 Tr 
Q_{Parisi}^2}\prod
\int_{-\infty}^{\infty}\frac{dz_{\gamma}^{(\alpha_{_{\gamma}})}}{\sqrt{2\pi}}
e^{-\frac{[z_{\gamma}^{^{(\alpha_{_{\gamma}})}}]^{^2}}{2}}
\prod\int d\bar{\psi}d\psi 
Exp[\sum_{\alpha_1=1}^{n/m}\sum_{a=(\alpha_1-1)m+1}^{\alpha_1m}
\nonumber\\
& &\sum_{i,\sigma,\epsilon_{_l}}(\bar{\psi}^{a,l}_{i,\sigma}
[{\cal{G}}_0^{-1}(\epsilon_l)+
\sigma\tilde{H}(\{z_{\gamma}^{(\alpha_{\gamma})}\})]
\psi^{a,l}_{i,\sigma}+
\eta_{i,\sigma}^{a,l}\bar{\psi}_{i,\sigma}^{a,l}-
\bar{\eta}_{i,\sigma}^{a,l}
\psi_{i,\sigma}^{a,l})] 
\end{eqnarray}
with a bare propagator ${\cal{G}}_0(\epsilon_l)=(i\epsilon_l+\mu)^{-1}$ 
and a magnetic field H included in the 
effective field $\tilde{H}(\{z_{\gamma}^{(\alpha_{\gamma})}\})= H + 
J\sum_{\gamma}
\sqrt{q_{\gamma}-q_{\gamma+1}} z_{\gamma}^{(\alpha_{\gamma})}$, where 
$q_{_0}\equiv\tilde{q}$, $q_{_{K+1}}=0$. Fermionic fields are denoted 
$\psi,\bar{\psi}$, and $\eta,\bar{\eta}$.
Spin (decoupling)--fields $z_{\gamma}$, carrying a Parisi block index,
explore the random magnetic order.
The Parisi matrix $Q_{Parisi}$ has the wellknown form 
\cite{parisi} apart from the nonvanishing diagonal elements 
$\tilde{q}$; 
their presence is required by the fact that $({\hat{\sigma}}^z)^2=
(\hat{n}_{\uparrow}-\hat{n}_{\downarrow})^2\neq 1$. 
The structure of the Parisi--matrix is of course responsible for 
the rather complicated form of the Lagrangian; despite this complication
the fermion fields can be eliminated in the standard way, which leads to 
the selfconsistent equations given below. \\
It is known since Parisi's work \cite{parisi} that an analytical 
low temperature expansion is hard to obtain even for the standard SK--model 
and its smaller set of selfconsistent parameters.
First insight is gained by the one--step RPSB ($K=1$).
The standard three parameter set of the SK--model for $K=1$, 
order parameters $q_1$ and $q_2$, and $m\equiv m_1\sim T$, 
is enlarged in the fermionic space by $\tilde{q}-q_1\sim T$, where
$\tilde{q}:=[<\sigma(\tau)\sigma(\tau^{\prime})>]_{av}$ represents
a spin correlation, which remains static unless a fermion hopping 
mechanism or other noncommuting parts are included in the Hamiltonian.
For the fermionic Ising spin glass the ($K=1$)--DoS reads 
\begin{equation}
\rho_{\sigma}(E)=\frac{ch(\beta\mu)+ch(\beta 
E)}{\sqrt{2\pi(\tilde{q}-q_1)}J}\frac{e^{-\frac{1}{2}\beta^2 
J^2(\tilde{q}-q_1)}}{\sqrt{2\pi q_2}}
\int_{-\infty}^{\infty}dv_2 
e^{-\frac{v_2^2}{2 q_2}}\frac{\int_{-\infty}^{\infty}dv_1
e^{-\frac{(v_1-v_2)^2}{2(q_1-q_2)}-\frac{(v_1+H+\sigma 
E)^2}{2(\tilde{q}-q_1)}}{\cal{C}}^{m-1}}{\int_{-\infty}^{\infty}dv_1 
e^{-\frac{(v_1-v_2)^2}{2(q_1-q_2)}}{\cal{C}}^m}
\end{equation}
with ${\cal{C}}=cosh(\beta \tilde{H})+\zeta$, where 
$\zeta=cosh(\beta\mu)exp(-\frac{1}{2}\beta^2(\tilde{q}-q_1))$
reveals the competition between the particle "pressure" exerted by the 
chemical potential $\mu$ and the single--valley susceptibility 
$\bar{\chi}=\beta(\tilde{q}-q_1)$ leading to a crossover 
at $|\mu|=\frac{1}{2}\bar{\chi}$ in the $T\rightarrow 0$--limit. 
The $\zeta$--term is a fermionic feature, absent from the standard 
SK--model. It is closely related to the fermion filling; this filling factor 
behaves discontinuously on the $T=0$--axis \cite{brro}.
For $T=0$ we obtain exactly
\begin{equation}
\rho_{\sigma}(E)=
\frac{e^{-\frac{1}{2}a^2(H)(1-q_2)-\frac{\Delta_E^2}{1-q_2}+a(H)\Delta_E-
\frac{H^2}{2 q_2}}}{\pi\sqrt{1-q_2(H)}}
\Theta(|E|-\bar{\chi})
\int_{-\infty}^{\infty}dz\frac{e^{-\frac{1}{2}\frac{z^2}{1-q_2}-
(\frac{\sqrt{q_2}}{1-q_2}\frac{\sigma E}{|E|}\Delta_E
-\frac{H}{\sqrt{q_2}})z}}{d(z)+d(-z)},
\label{4}
\end{equation}
\begin{equation}
d(z)\equiv e^{a(H)\sqrt{q_2}z}\left[1+Erf(\frac{a(H)(1-q_2)+
\sqrt{q_2}z}{\sqrt{2(1-q_2)}})\right],\quad \Delta_E\equiv |E|-\bar{\chi},
\end{equation}
and $a(H)\equiv m'(T=0)$. The replica--symmetric solution
displays a magnetic hardgap of width $2 E_g(H)$ in the DoS and the system
remains half--filled at $T=0$ within the finite interval given by 
$|\mu|<\frac{1}{2}E_g(H)$. For higher values of the chemical potential,
hence smaller spin density, phase separation occurs together with a 
discontinuous transition into a full or an empty system \cite{brro}. 
A stable homogeneous saddle--point solution could only be found for the 
half--filled case. Thus the following analysis is 
restricted to this interval of chemical potentials. Its width is determined 
selfconsistently and seen to decrease to zero as $K\rightarrow0$.
The selfconsistent equations for $\tilde{q}, q_1, q_2$ and 
the Parisi parameter $m$ \cite{parisi} simplify in the $T=0$--limit becoming
\begin{eqnarray}
& &\hspace{-.5cm}\tilde{q}=q_1=1,\hspace{.2cm} 
lim_{T\rightarrow0}\frac{\tilde{q}-q_1}{T}=\bar{\chi},\hspace{.2cm}
q_2=\int_{-\infty}^{\infty}\frac{dz}{\sqrt{2\pi}}
e^{-\frac{(z-H/\sqrt{q_2})^2}{2}}\left[\frac{d(z)-
d(-z)}{d(z)+d(-z)}\right]^2
\\
& &\hspace{-.5cm}0=1-q_2^2-\frac{4}{a}
\int_{-\infty}^{\infty}\frac{dz}{\sqrt{2\pi 
}}\hspace{.1cm}e^{-\frac{(H/\sqrt{q_2}-z)^2}{2}}
\{-\frac{1}{a}\hspace{.1cm}ln[\frac{1}{2}e^{\frac{1}{2}a^2 t} 
(d(z)+d(-z))]\nonumber\\ 
& &\hspace{-.3cm}+[(at+\sqrt{q_2}\hspace{.1cm}z)d(z)+
(at-\sqrt{q_2}\hspace{.1cm}z)d(-z)+\sqrt{8 
t/\pi}\hspace{.1cm}e^{-\frac{1}{2}a^2 
t-\frac{1}{2}q_2 z^2/t}] /[d(z)+d(-z)]\} 
\end{eqnarray}
where $t\equiv q_1-q_2$.
For zero magnetic field one finds
$q_2=0.476875, a=lim_{T\rightarrow 0}m'(T=0)=1.36104$, and
$\bar{\chi}=lim_{T\rightarrow 0}(\tilde{q}-q_1)/T=.239449$. 
The H--dependent solutions shown in fig.1 are obtained numerically and 
then used in evaluating eq.(\ref{4}) for the density of states. 
$T=0$--results are shown in figs.2 and 3, while
the result at finite low temperature of fig.4 illustrates the presence of 
plateaus of constant slope, each corresponding to Parisi order parameter
separations (here: $\tilde{q}-q_1$ and $q_1-q_2$). The number of these 
plateaus of constant slope increases with the order K of Parisi--RPSB.
Hence, the time--dependence of the Green's function should 
characteristically depend on the order parameter separations 
${q_k-q_{k-1}}$.
If we compare with the replica--symmetric result a reduction of 
the gapwidth is observed. Analytically one finds a gapwidth
\begin{equation}
E_g (H)=\bar{\chi}=lim_{T\rightarrow 0}\beta(\tilde{q}-q_1)
\end{equation}
which turns into $lim_{T\rightarrow 0}(\beta(\tilde{q}-q(1)))$ 
in terms of the Parisi function at $K=\infty$.
Only at zero RPSB this susceptibility coincides with the equilibrium $\chi$.
In fact for 1--step RPSB the fermionic Ising spin glass approaches
$\chi=\beta(\tilde{q}-q_1)+\beta m(q_1-q_2)\rightarrow .95$,
the same numerical value as the one for the SK--model.
\begin{figure}
\epsfig{file=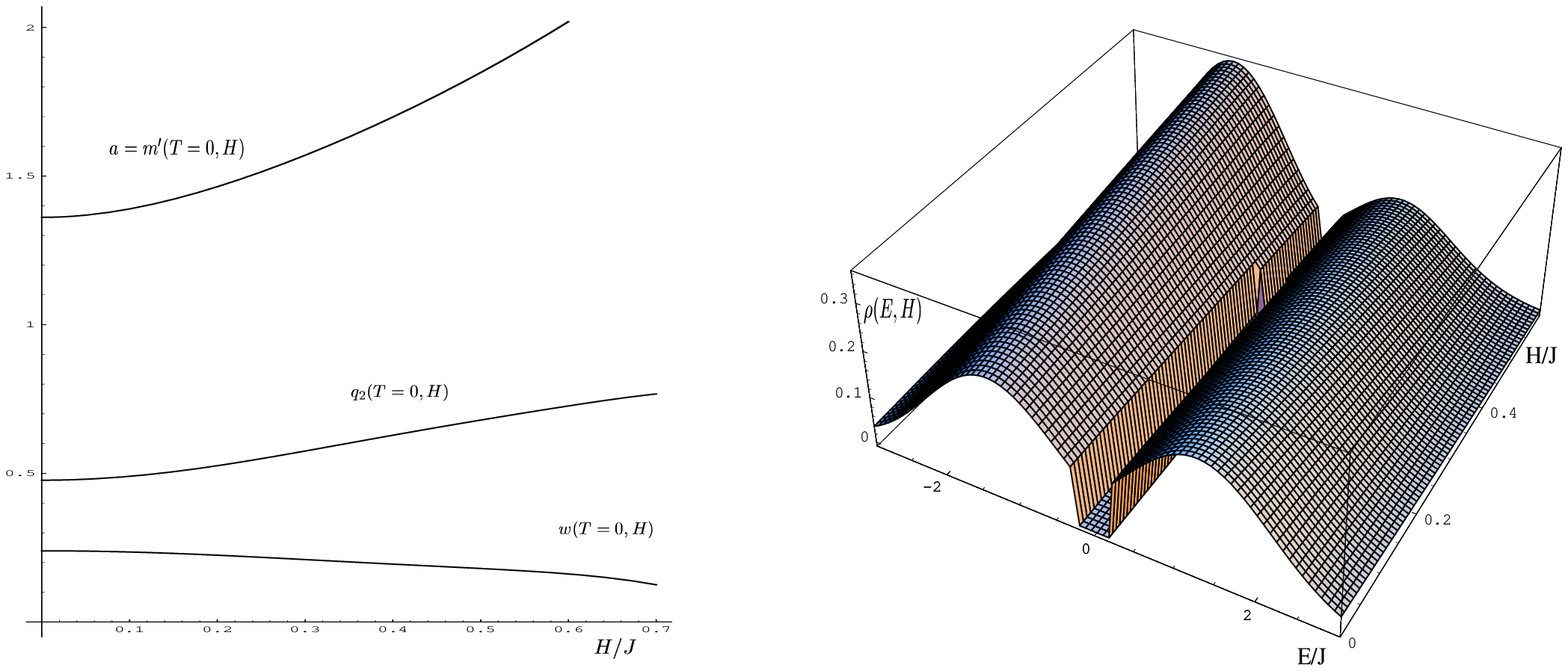,width=5.7cm,angle=270}
\caption{Field dependence of $dm/dT$ (top),
of the order parameter $q_2$, and of gapwidth parameter w (bottom) for 1RPSB
and zero temperature.}
\caption{density of states at $T=0$ as a function of energy and
magnetic field for 1--step RPSB}
\end{figure}
\begin{figure}
\epsfig{file=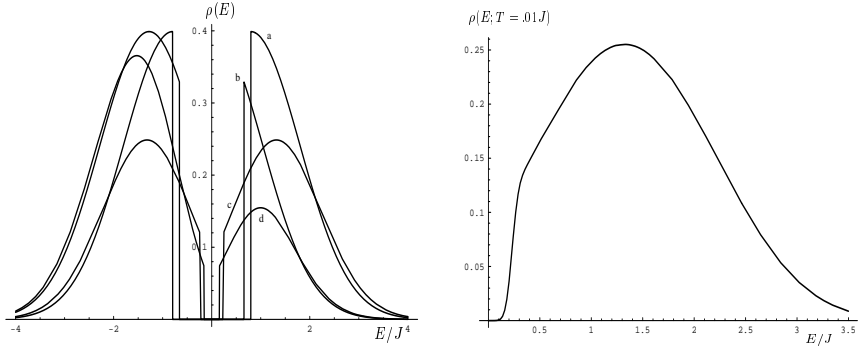,width=5cm,angle=270}
\caption{Effect of one step replica symmetry breaking on the fermionic
density of states (DoS) for magnetic fields $H=0$ (curve c: 1RPSB, a: 0RPSB)
and $H/J=0.6$ (d: 1RPSB, b: 0RPSB)}
\caption{Low but finite temperature ($T=.01 J$) form of the zero
field DoS in 1--step RPSB.}
\end{figure}  
While the 1--step RPSB provides already a much better approximation 
than 0--RPSB it is still unstable towards higher RPSB.
We have therefore extended given equations to arbitrary K.
Apart from the K-invariant relation $E_g(H)=\bar{\chi}$ we find a second
invariant with respect to K--th RPSB, including $K=\infty$, which is given 
by
\begin{equation}
lim_{_{|E|\downarrow E_g(H)}}\rho_{\sigma}(E)=\frac{1}{2} 
\bar{\chi}=\frac{1}{2}E_g(H)
\label{10}
\end{equation}
The invariant ratio $1/2$ is seen analytically by comparing the formulae for 
the gapwidth and for $\rho(|E|=E_g+0)$ (both generalized to arbitrary K) 
in the $T\rightarrow 0$--limit. For each given K, (half) the gapwidth 
equals the fermionic nonequilibrium susceptibility $\bar{\chi}$, which 
turns into $\bar{\chi}=\beta(\tilde{q}-q(1))$ differing only by 
exponentially small terms from the SK--model result
$\bar{\chi}=\beta(1-q(1))\sim T$ \cite{binyg}, 
where $q(1)$ denotes the Parisi function $q(x)$ at $x=1$. 
Consequently the DoS--hardgaps at finite K terminate in a softgap
for $K\rightarrow \infty$. Note that we did not have to evaluate the 
$T=0$--Parisi function $q(x)$ in order to reach this conclusion.
Assuming that the relation between gapwidth and $\bar{\chi}$ remains valid
(at least in good approximation) for short--range models, fluctuation effects
should harden the gap. This requires further analysis.\\
A quantity of particular interest in many--body theories is the
Ward identity for charge conservation. It shows that Parisi symmetry 
breaking, as observed in the density of states, exists also in
vertex functions. The Ward identity for the insulating model can be 
viewed as one of a metallic spin glass at momentum 
transfer $\underline{k}=0$, ie
$i\omega\hspace{.1cm}\Lambda_{n}(\underline{k}=0,\epsilon+\omega,\epsilon)=
{\cal{G}}(\epsilon+\omega)-{\cal{G}}(\epsilon)$ (fermion momenta 
suppressed). 
$\Lambda_{n}$ is the Fourier transformed three point function
$<T_{\tau}[a_i^{\dagger}(\tau)a_{i^{\prime}}(\tau')\hat{n}_j(0)]>$ in 
terms 
of the 
fermion operators. Thus the density--part $\Lambda_n$ in the Ward 
identity (a current--part $\Lambda_{\underline{j}}$ emerges for 
itinerant models and $\underline{k}\neq 0$) obeys
\begin{equation}
lim_{\omega\rightarrow 
0}\hspace{.1cm}\omega\hspace{.1cm}lim_{\underline{k}\rightarrow 0}
\Lambda_n^{AR}(\underline{k},\epsilon+\omega,\epsilon)=
2\pi i\hspace{.1cm}\rho(\epsilon,\{q_r-q_{r-1}\})\nonumber
\end{equation}
and thus shows that the Parisi form of the DoS, depending on all 
$q_r-q_{r-1}$ or on $q(x)$ for $K=\infty$, also enters the vertex function. 
This will also occur in metallic spin glases, whence diffusive modes and
conductivity are expected to depend on Parisi symmetry breaking.
While we have proved the existence of a spin--glass hardgap at any finite 
$K>0$ with
\begin{equation}
\delta\rho(E)\sim |E-w(H)|\quad, [|E|\geq \bar{\chi}, K<\infty];\quad\quad
\rho(E)\sim |E|^x, [K=\infty]
\end{equation}
the pseudogap solution at $K=\infty$ has a scaling exponent x, which could 
eventually become different from one and remains to be determined.
This pseudogap together with x=1 would be slightly reminiscent of the exponent 
found for a superconducting glass unitary nonlinear sigma model \cite{ro}.
We remark that the pseudogap--solution given for the fermionic spin glass 
model refers precisely to $\mu=0$ whereas the hardgap--solutions at 
finite K happened to be stable within finite intervals $|\mu|\leq 
\bar{\chi}/2$ corresponding to half--filling only at $T=0$. 
The regime beyond half--filling, identified as the domain of phase 
separation in the replica--symmetric solution \cite{brro}, requires 
further analysis at $T=0$ as well as several metallic--, Kondo--type--, 
superconducting--, and other model extensions. The new method of Fourier 
Transformations in replica space \cite{temesvari} is hoped to facilitate 
further insight into the difficult $K=\infty$--solutions.\\
We note that an overlap distribution function for data clustering 
shown in \cite{LootensBroeck} and interpreted as a pseudo $T=0$--problem in 
a classical spin analogy, revealed, apart from the ratio discussed in 
Eq.(\ref{10}), a remarkable similarity with the $(H=0)$--density of states. 
It appears interesting to explore pseudo--$(T=0)$ neural network 
problems \cite{gyrei} as potential classical partners of fermionic 
spin glasses.\\
Summarizing our results we proved i) replica permutation symmetry 
breaking to be most important in the $T=0$ quantum field theory of the 
fermionic spin glass, 
ii) that low energy excitations are determined 
by RPSB and hence the long--time quantum--dynamical behaviour of the fermion 
Green's function carries RPSB--fingerprints, which iii) affects higher
order correlations by means of charge conservation too.\\ 
This work was supported by the DFG under project Op28/5.
One of us (R.O.) wishes to thank for hospitality extended to him at the ICTP 
Trieste (Italy) 
during the workshop on {\it Statistical physics of frustrated systems}, 
where part of this work was carried out, and for valuable remarks by 
John Hertz, who draw our attention to ref.\cite{gyrei}, by Giorgio Parisi, and
David Sherrington.

\end{document}